\documentclass[aps,superscriptaddress,twocolumn,showpacs,amsmath,amssymb]{revtex4-1}

\usepackage{slashed}
\usepackage{graphicx}
\usepackage{color}
\usepackage{wrapfig}
\usepackage{braket}

\newcommand{\bea}{\begin{eqnarray}}
\newcommand{\eea}{\end{eqnarray}}
\newcommand{\beq}{\begin{equation}}
\newcommand{\eeq}{\end{equation}}

\begin{document}
\title{Canonical approach to finite density QCD with multiple  precision 
computation}

\author{Ryutaro Fukuda}
\affiliation{Department of Physics, The University of Tokyo,7-3-1 Hongo, Bunkyo-ku, Tokyo 113-0033, Japan}
\affiliation{Institute f\"{u}r Theoretische Physik, ETH Z\"{u}rich, CH-8093 Z\"{u}rich, Switzerland}

\author{Atsushi Nakamura}
\affiliation{Research Center for Nuclear Physics, Osaka University, Ibaraki 567-0047, Japan}

\author{Shotaro Oka}
\affiliation{Department of Physics, Rikkyo University, 3-34-1 Nishi-Ikebukuro, Toshima-ku, Tokyo 171-8501, Japan}

\begin{abstract}
%In this work, we mainly discuss the following two topics.
%One is the validity of the canonical approach to finite density QCD 
%by contrast with multi parameter reweighting method.
%The other is the baryon chemical potential dependence of thermodynamic observavles 
%such as pressure, baryon number density and susceptibility.
%We make clear that the canonical approach is consistent with multi parameter reweighting method
%and our results are reliable over $\mu_ B/T=3$ in most cases. 
%In addition, we refer that multiple length precision calculation plays an important roll
%in the evaluation of canonical partition functions.
%
We calculate the baryon chemical potential ($\mu_B$) dependence 
of thermodynamic observables, 
i.e., pressure, baryon number density and susceptibility
by lattice QCD using the canonical approach. 
We compare the results with those by the multi parameter reweighting (MPR)
method;
Both methods give very consistent values in the regions where errors of
the MPR are under control.
The canonical method gives reliable results over $\mu_ B/T=3$,
with $T$ being temperature. 
Multiple precision operations play an important roll
in the evaluation of canonical partition functions.

\end{abstract}
\maketitle

%%%%%%%%%%   Introduction   %%%%%%%%%%
\section{Introduction}

Quantum Chromodynamics (QCD) is the fundamental theory describing the strong interaction.
It is well known that QCD has the rich phase structure at finite temperature and density
\cite{Fukushima:2011jc}.
And yet, regions that we can access with the  perturbation are limited.
Currently, the most promising method to explore the phase diagram is lattice QCD simulation 
which is first principle calculation of QCD. 

Although the lattice QCD simulations are very successful to 
analyze the phase diagram of a finite temperature system, at finite density
they have a severe problem, so-called sign problem, 
and outcomes of the first-principle calculation can be available only 
at small chemical potential range.
In finite temperature and density systems, lots of  physically 
interesting targets such as the early universe, neutron stars, 
quark matters are waited to be explored. 
Therefore, it is quite desirable to explore methods for investigating finite 
density QCD systems from ab initio calculation; this is the one of the 
urgent subjects in particle physics and nuclear physics.

The canonical approach we study in this paper is a promising candidate 
for this purpose.
%In Ref.\cite{Nakamura:2015yga}, the chiral condensate as well as the 
%thermodynamic quantities is studied in the canonical approach in a wide
%range of the temperature and chemical potential regions, and an indication
%of the transition was observed below $T_c$ and finite baryon density.
In Ref.\cite{Nakamura:2015yga}, the fugacity expansion by a method of
the hopping parameter expansion was constructed as a winding number expansion,
and the chiral condensate as well as the thermodynamic quantities is studied.
More detailed analyses were performed in Ref.\cite{Nakamura:2015qcd}
in a wide range of the temperature and chemical potential regions,
and an indication of the transition was first observed below $T_c$
and finite baryon density.
In this paper, we address two questions:
\begin{enumerate}
\item
Does the lattice canonical approach produce consistent results with the MPR ?
\item
In obtaining the canonical partition functions for large baryon number,
what is a role of the multi precision calculations ?
\end{enumerate}

\subsection*{Basic concept of the canonical approach in QCD}

In $N_f$ flavor QCD case with the degenerate quark masses,
the grand canonical partition function at finite temperature $T$ and
finite quark chemical potential $\mu_q$ is given in the path integral formalism as follows.
\begin{equation}
Z_{GC}(T,\mu_q)=\int d[U]\{\det\Delta(\mu_q)\}^{N_f}e^{-S_\mathrm{g}},
\end{equation}
where $\det\Delta(\mu_q)$ is the one flavor fermion determinant and $S_\mathrm{g}$ is the gauge action.
Because the fermion determinant has the property
\begin{equation}
\label{det}
[\det\Delta(\mu_q)]^\ast=\det\Delta(-\mu_q^\ast),
\end{equation}
the Monte Carlo measure $\{\det\Delta(\mu_q)\}^{N_f}e^{-S_\mathrm{g}}$ becomes complex number 
at finite real chemical potential and the standard Monte Carlo method breaks down. 
Consequently, we cannot study finite density thermodynamics with standard grand canonical method.
This difficulty is called sign problem.

A system described by the grand canonical partition function 
$Z_{GC}(T,\mu_q)$ is equivalent to a system described by the canonical partition function $Z_{C}(n,T)$ with fugacity
 $e^{\mu_q/T}$ in thermodynamic limit.
The relation of two ensembles can be written as a fugacity expansion\cite{FugacityExpansion}
using eigen vectors of number operator $\hat{N}\ket{n}=n\ket{n}$,

 \begin{align}
 \label{canonical}
 Z_{GC}(T,\mu_q) &= 
 \mathrm{Tr}\hspace{1mm}e^{-(\hat{H}-\mu_q\hat{N})/T} \notag \\ 
 &=\sum_{n=-\infty}^\infty\bra{n}e^{-\hat{H}/T}\ket{n}e^{n\mu_q/T} \notag \\ 
 &\equiv\sum_{n=-\infty}^\infty Z_C(n,T)\hspace{1mm}e^{n\mu_q/T},
 \end{align}
where $e^{\mu_q/T}$ is fugacity.
If we have the canonical partition functions, $Z_C(n,T)$,  for all net quark numbers $n$,
we can construct the grand canonical partition function as a polynomial 
of fugacity with coefficients $Z_C$.
From this formula, one can obtain Lee-Yang zeros\cite{Nakamura-Nagata13}, 
which reflect the system's critical nature\cite{Yang:1952be}.
 
The canonical partition functions are constructed through
the Fourier transformation of grand canonical partition function
at pure imaginary chemical potential\cite{Hasenfratz-Toussaint},
 \begin{equation}
\label{fourier}
Z_C(n,T)=\frac{1}{2\pi}\int_0^{2\pi}d\left(\frac{\mu_I}{T}\right)Z_{GC}\left(\frac{i\mu_I}{T}\right)e^{-in\mu_I/T},
\end{equation}
where $\mu_I\in\mathbb{R}$.
Eq.(\ref{det}) tells us that
the fermion determinant is real in the case of pure imaginary chemical potential.
Monte Carlo simulations can then be performed 
and the canonical partition functions are obtained by Eq.(\ref{fourier}).
Eq.(\ref{fourier}) also insists that the canonical partition fuctions are real number
because the grand canonical partition function 
is even function (charge conjugation invariant) in terms of chemical potential.
Considering this feature with Eq.(\ref{canonical}), one can find that canonical partition functions are
real and positive also in the context of canonical approach.

Once $Z_C$ are available, we can construct the grand partition function
by Eq.(\ref{canonical})
at any \textit{real} quark chemical potential.
This is because the chemical potential dependence of the grand canonical 
partition function appears only through fugacity, $e^{\mu_q/T}$, which is the variable
of the polynomial, and not in the coefficients, $Z_C$ in Eq.(\ref{canonical});
the effect of the chemical potential appears through the fugacity and 
the canonical partition function plays just a role of coefficients 
in the fugacity expansion of the grand canonical partition function.

\section{Framework}

\subsection{Winding number expansion of grand Partition function}

In this work, we employ the RG-improved gauge action
\begin{equation}
S_{\mathrm{g}}=\frac{\beta}{6}\left[c_0\sum_{n,\mu<\nu}W_{\mu\nu}^{1\times1}(n)
+c_1\sum_{n,\mu<\nu}W_{\mu\nu}^{1\times2}(n)\right]
\end{equation}
with $c_1=-0.331$ and $c_0=1-8c_1$,
and the clover improved Wilson fermion action with the quark matrix
\begin{align}
\label{wilson}
 \Delta(n,m,\mu_q)=&\delta_{nm}
-\kappa C_{SW}\delta_{nm}\sum_{\mu\leq\nu}\sigma_{\mu\nu}F_{\mu\nu} \notag \\
&-\kappa\sum_{i=1}^3\Big[\left(1-\gamma_i\right)U_i(n)\delta_{m,n+\hat{i}} \notag \\
&\hspace{25mm}+\left(1+\gamma_i\right)U_i^\dagger(m)\delta_{m,n-\hat{i}}\Big] \notag \\
&-\kappa\Big[e^{+\mu_qa}(1-\gamma_4)U_4(n)\delta_{m,n+\hat{4}} \notag \\
&\hspace{17.5mm}+e^{-\mu_qa}(1+\gamma_4)U_4^\dagger(m)\delta_{m,n-\hat{4}}\Big]  \notag \\
&\equiv1-\kappa Q(\mu_q).
\end{align}
Here $n,m$ are space-time coordinates on a lattice, $\kappa$ is hopping parameter and $\mu_q$ is the quark chemical potential which is introduced to the temporal part of link variables.

In order to obtain canonical partition functions,
we need to compute the grand canonical partition functions
at various pure imaginary chemical potential values in Fourier 
transformation Eq.(\ref{fourier}).

We use the reweighting method to evaluate the grand canonical partition function, 
\begin{align}
\label{Zapi}
Z_{GC}(i\mu_I)&=
\int dU\left[\frac{\det\Delta(i\mu_I)}{\det\Delta(\mu_0)}\right]^{N_f}
\left\{\det\Delta(\mu_0)\right\}^{N_f}\hspace{1mm}e^{-S_{g}} \notag \\
&=\left<\left[\frac{\det\Delta(i\mu_I)}{\det\Delta(\mu_0)}\right]^{N_f}\right>_{\mu_0}Z_{GC}(\mu_0),
\end{align}
where
$\mu_0 = 0$ or pure imaginary values.
We can then evaluate the canonical partition function as
\begin{equation}
\label{const}
\frac{Z_C(n,T)}{Z_C(0,T)}=\frac{1}{2\pi}\int_0^{2\pi}d\left(\frac{\mu_I}{T}\right)
Z_{GC}(i\mu_I)\hspace{1mm}e^{in\mu/T}.
\end{equation}
We adopt, here, normalized canonical partition functions Eq.(\ref{const}) 
in order to avoid the extra constant $Z_{GC}(\mu_0)$ in Eq.(\ref{Zapi});
this step does not affect the physical result.
Now the evaluation of the grand canonical partition function is reduced to
the calculation of  ratios of fermion determinants in Eq.(\ref{Zapi}).

Performing the hopping parameter expansion in the logarithm,
we write the fermion determinant as 
\begin{eqnarray}
\label{hpe}
\det\Delta(i\mu_I)=\exp\Big[\mathrm{Tr}\log\{1-\kappa\Delta(i\mu_I)\}\Big]
\nonumber \\
=\exp\Bigg[-\mathrm{Tr}\sum_{j=1}^\infty
\frac{\kappa^i}{j}Q^j(i\mu_I)\Bigg].
\end{eqnarray}
The trace is taken over space-time, spinor and color. 
Here, we used the following identity for arbitrary matrix A:
\begin{equation}
\det A = e^{\log\det A}=e^{\mathrm{Tr}\log A},
\end{equation}
and $\log$ is expanded assuming $\kappa$ is small (hopping parameter expansion).

The contribution of the trace in Eq.(\ref{hpe}) comes from all closed loops on a lattice,
and
the chemical potential dependence comes from specific closed loops 
winding along positive and negative time directions.
We can thus classify the trace of the hopping parameter expansion 
in Eq.(\ref{hpe})
according to the winding number which is the number of net windings along the time direction.
As a result, we can reach following expression with coefficients $W_n$ 
and complex fugacity $e^{i\mu_I/T}$.
Here $n$ represents the winding number.
\begin{equation}
\label{wne}
\det\Delta(i\mu_I)=\exp\left[\sum_{n=-\infty}^\infty W_n e^{in\mu_I/T}\right].
\end{equation}
We call this expression as `winding number expansion'.
The negative winding number appeared in Eq.(\ref{wne}) stands for
the winding along negative time direction. 
The coefficients $W_n$ has no chemical potential dependence;
the chemical potential dependence appears in the fugacity.
Consequently, we have only to calculate $W_n$ from given gauge configurations 
to obtain grand canonical partition functions
at desired pure imaginary chemical potential.
  
 \subsection{Constraint on canonical partition function from symmetry of QCD}

Roberge and Weiss pointed out that the QCD grand canonical partition 
at pure imaginary chemical potential has the following periodicity\cite{Roberge-Weiss}
\begin{equation}
\label{rws}
Z_{GC}\left(\frac{i\mu_I}{T}\right)=Z_{GC}\left(\frac{i\mu_I}{T}+\frac{2\pi ik}{3}\right),
\end{equation}
where $k\in\mathbb{N}$.
Using Eq.(\ref{rws}), we rewrite the grand canonical partition function as 
\begin{equation}
Z_{GC}\left(\frac{i\mu_I}{T}\right)=
\frac{1}{3}\sum_{k=0}^2Z_{GC}\left(\frac{i\mu_I}{T}+\frac{2\pi ik}{3}\right) .
\end{equation}
Then, we get the following relation,
\begin{align}
\label{rotation}
Z_C(n,T)=\frac{1}{2\pi}\int_0^{2\pi}&d\left(\frac{\mu_I}{T}\right)Z_{GC}\left(\frac{i\mu_I}{T}\right)e^{-in\mu_I/T} \notag \\
&\hspace{7mm}\times\left[\frac{1+e^{i\frac{2\pi}{3}n}+e^{i\frac{4\pi}{3}n}}{3}\right] .
\end{align}
We obtain the following important constraint on the canonical partition functions,
\begin{equation}
Z_C(n\ne 3k)=0.
\end{equation}
Note that this holds both in the confinement and the deconfinement phases.

Now the grand partition function can be written as
\begin{equation}
Z_{GC}(T,\mu_B)=\sum_{B=-\infty}^{\infty}Z_{C}(B,T)e^{B\mu_B/T},
\end{equation}
where $B\in\mathbb{N}$.
Because this quantum number $B$ can be interpreted as net baryon number, 
$\mu_B$ can be regarded as baryon chemical potential
which is related to quark chemical potential as $\mu_B=3\mu_q$.

\subsection{Thermodynamic observables}

In a homogeneous system, the dimensionless equation of state at $(\mu_B,T)$ is given by 
\begin{align}
\frac{p(\mu_B,T)}{T^4}&=\frac{1}{V_sT^3}\log Z_{GC}(\mu_B,T) \notag \\
&=\left(\frac{N_t}{N_s}\right)^3\log Z_{GC}(\mu_B,T),
\end{align}
where $N_s=N_x=N_y=N_z$ and $T^{-1}=N_ta$ with a lattice spacing $a$.
The deviation of the pressure from $\mu_B=0$ is given by 
\begin{align}
\frac{\Delta p(\mu_B,T)}{T^4}&=\frac{p(\mu_B,T)}{T^4}-\frac{p(0,T)}{T^4} \notag \\
&=\left(\frac{N_t}{N_s}\right)^3 
\log\left(\frac{Z_{GC}(\mu_B,T)}{Z_{GC}(0,T)}\right) .
\end{align}

The dimensionless baryon number density $n_B/T^3$ and susceptibility $\chi/T^2$ are
\begin{equation}
\frac{n_B(\mu_B,T)}{T^3}=\frac{\partial}{\partial(\mu_B/T)}\frac{p(\mu_B,T)}{T^4} ,
\end{equation}
\begin{equation}
\frac{\chi(\mu_B,T)}{T^2}=\frac{\partial^2}{\partial(\mu_B/T)^2}\frac{p(\mu_B,T)}{T^4}.
\end{equation}

\section{Numerical results}

\subsection{Lattice set up}
We adopt $2$-flavor clover improved Wilson fermion action 
with $C_{\mathrm{SW}}=(1-0.8412/\beta)^{-3/4}$ evaluated by
one-loop perturbation theory and RG-improved gauge action.
All simulations were performed on $ N_x\times N_y\times N_z\times N_t=8\times8\times8\times4$ lattice.
We considered $\beta$ = 2.00, 1.95, 1.90, 1.85, 1.80, 1.70
which correspond to 
$T/T_c$ = 1.35(7), 1.20(6), 1.08(5), 0.99(5), 0.93(5), 0.84(4)\cite{WHot2010}.
The values of hopping parameter $\kappa$ was determined for each $\beta$ by following the line
of the consistent physics in case of $m_\pi/m_\rho=0.8$ in Ref.\cite{WHot2010}.

We generated gauge configurations at $\mu_0=0$ with the hybrid Monte Carlo (HMC) method.
The step size $d\tau$ and number of steps $N_\tau$ of HMC were set to $\delta \tau=0.02$,
$N_\tau=50$ so that the simulation time was $d\tau \times N_\tau=1$.
After the first $2000$ trajectories for thermalization, we adopted  $400$ configurations every 200 trajectories
for each parameter set.

\subsection{Instability of Fourier transformation in canonical approach and its solution}

Before proceeding to our numerical results, we refer to the instability of Fourier transformation in canonical approach
and then discuss our strategy to avoid it in this subsection.

Since the fugacity expansion of grand canonical partition function, Eq.(\ref{canonical}),
converges at real baryon chemical potential, 
the canonical partition function $Z_n$ must become smaller when the net baryon 
number $|n|$ becomes large. 
This means that we have to deal with quite small values as the results of Fourier transformation.
This step is quite difficult in the point of view of numerical calculation
because the Fourier transformation is an oscillatory integral.

\subsubsection{Instability of Fourier transformation}
\begin{figure}
	\center
	\includegraphics[width=1\hsize, bb=0 0 435 244, clip]{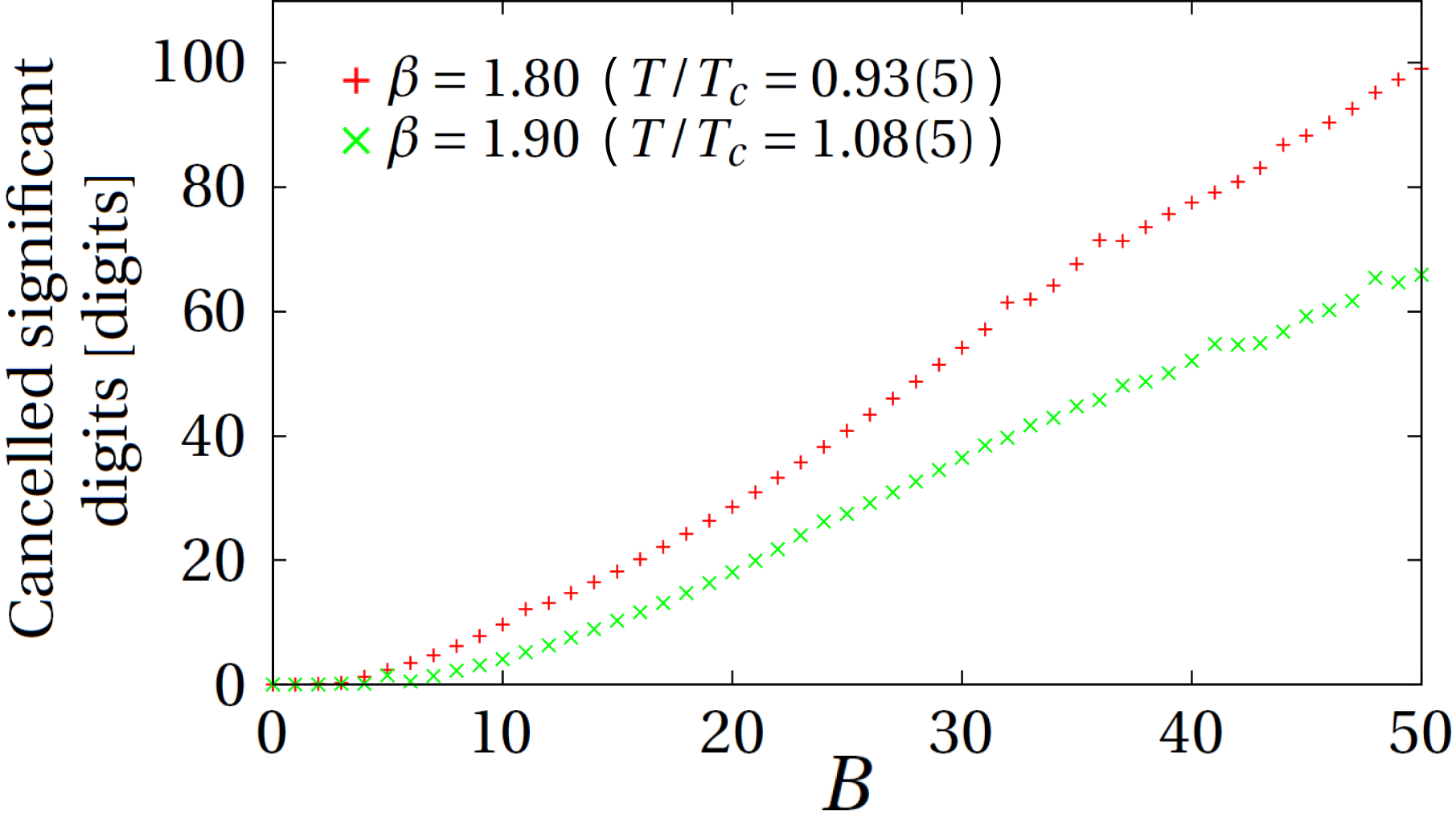}
	\caption{\label{drop} (Color online) Cancelled significant digits in calculations of DFT 
				at over $T_c$ (upper red points) and below $T_c$ (lower green points).}
\end{figure}

In numerical calculation, Fourier transformation Eq.(\ref{fourier}) is 
computed by discrete Fourier transformation (DFT) as
\begin{equation}
	Z_{C} (n, T) = \frac{1}{N} \sum_{k=0}^{N-1} Z_{GC} \left(i \frac{\mu_I}{T} = i\frac{2 \pi  k}{N} \right) e^{i \frac{2 \pi k}{N} n},   \label{DFT}
\end{equation}
where $N$ is the interval number of DFT.
Because DFT is just a discretized version of Fourier transformation in continuum theory,
the instability of DFT in canonical approach is simply
caused by the numerical errors.
They are classified into
rounding error, truncation error, cancellation of significant digits and loss of trailing digits.
The instability of DFT does not come from truncation error since DFT is not infinite series.
Accordingly, it is quite natural to consider that the instability originates from cancellation of significant digits 
or loss of trailing digits or both of them.
In this work, we actually monitored the behavior of all variables in our DFT program in order to study the effect of these two errors.
As the result, we found that cancellation of significant digits is not negligible in DFT program.
Fig.\ref{drop} represents cancelled digits in DFT.
For example, 80 digits are cancelled in case of $\beta=1.80$, $B = n / 3 = 40$.
We also found that the appearance of the cancellation does not depend on temperature of a system.

\subsubsection{Solution of instability - multiple length precision calculation}

Cancellation of significant digits arises from the following types of calculation,
\begin{equation}
	\begin{array}{lll}
		1.234567 - 1.234566 & = & 0.000001  \\
		\textrm{(7 significant digits)} & \phantom{= } & \textrm{(1 significant digit)}
	\end{array}
\end{equation}
In this case, six significant digits are lost.

In order to reduce the effect of this cancellation, 
we should increase significant digits as a solution.
Let us consider the next calculation with 22 significant digits.
\begin{align}
	&\begin{array}{l}
		1.234567444444444444444 - 1.234566111111111111111  \\
		\textrm{(22 significant digits)}
	\end{array} \notag \\
&\hspace{25mm}
	\begin{array}{ll}
		= 	& 0.000000133333333333333 \\
			& \textrm{(16 significant digits)}
	\end{array}
\end{align}
Although six significant digits are definitely lost in this calculation,
16 significant digits survive in the final result.

\begin{figure*}
	\center
	\includegraphics[width=0.8\hsize, bb=0 0 744 256, clip]{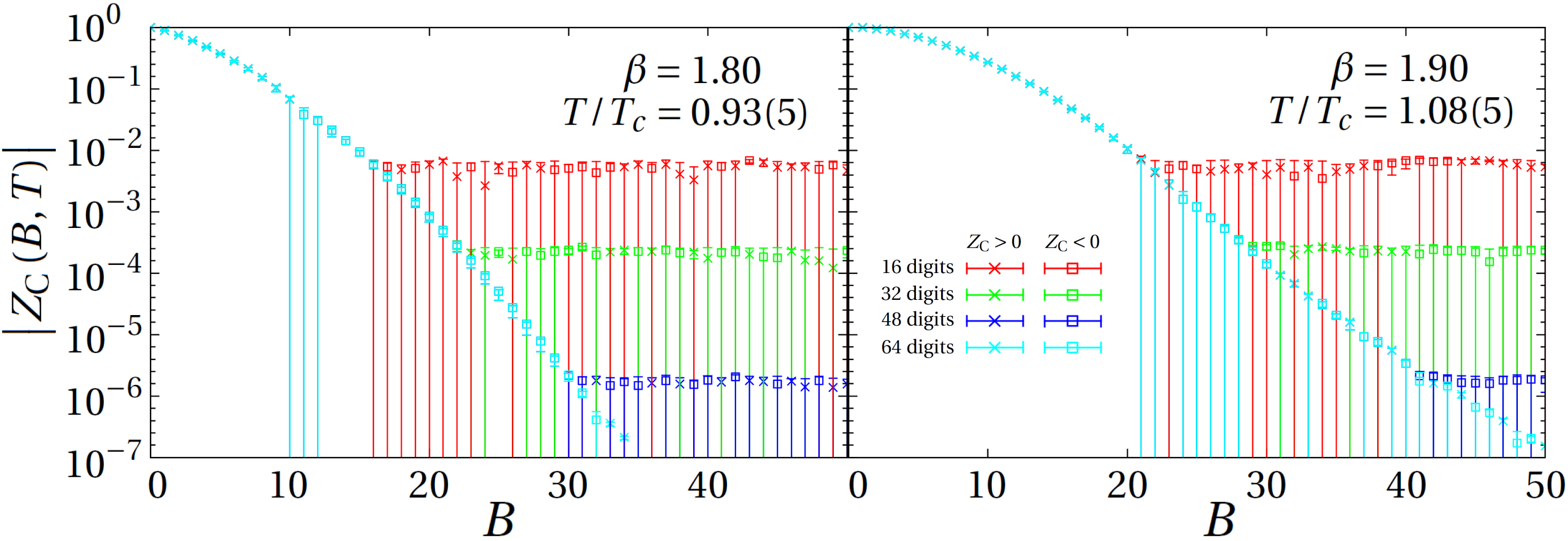}
	\caption{\label{Zndrop} (Color online) Relation between behavior of $Z_{C} (B,T)$ and 
			precision of variables, at over $T_c$ (left panel) and below $T_c$ (right panel).
			Both panels are plotted with some precision 
			at 16 digits (double precision, uper red points), 32 digits (second green points),
			48 digits (third blue points), and 64 digits (lowest cyan points).}
\end{figure*}

Summarizing above, the precision of the result can be kept by increasing significant digits of variables in this way.
Fig.\ref{Zndrop} represents the cancelled digits in the calculation of $Z_{C} (B, T)$ 
with 16, 32, 48, 64 precision calculation.
According to this figure, 
for evaluating $Z_{C} (B, T)$ to larger $n$,
it is essential to increase significant digits of variables.

\subsection{Thermodynamic observables at finite real baryon chemical potential}
\subsubsection{Calculation procedure}

First, we computed coefficients of winding number expansion $W_n$ up to $n=120$ with $400$ configurations 
in all temperature cases.
We used $64$ (above $T_c$) and $128$ (below $T_c$) noise vectors to calculate
the trace in the fermion determinant Eq.(\ref{hpe}). 
Then, we evaluated the grand canonical partition functions at various pure imaginary chemical potentials
using the winding number expansion with $W_n$.
After that, we evaluated the canonical partition function through Fourier transformation
and thermodynamic observables.
We adopted multiple length precision calculation\cite{FMlib}  with $400$ significant digits
in order to keep the sufficient precision except for the calculation of gauge configurations and $W_n$.
Gauge configurations and $W_n$ were computed with double precision, 
i.e., around 16 significant digits.

Note that canonical partition functions are complex number in numerical calculations
because of numerical errors.
Therefore, we adopt the only real part of the canonical partiton function. 
If the real part of the canonical partition function is negative at some baryon number $n_B$
fo the first time, we adopt the result up to $n_B-1$ as canonical partiton functions.

\subsubsection{Estimation of truncation error in fugacity expansion}
In numerical calculation, we have to deal with 
the fugacity expansion of the grand canonical partition function as the finite series
\begin{equation}
\label{trun}
Z_{GC}(T,\mu_B)=\sum_{B=-N_{\mathrm{max}}}^{N_{\mathrm{max}}}Z_C(B,T)e^{B\mu_B/T}.
\end{equation}
Therefore, we have to judge the baryon chemical potential region 
where results are free from the truncation error.
There may be several possible ways to analyze the effect of the truncation 
error;
In this work,
we use the following analysis. 

First, we evaluate the expectation values of thermodynamic observables $\left<O(\mu_B)\right>_{N_{\mathrm{max}}}$
with Eq.(\ref{trun}).
Next, we calculate the expectation values $\left<O(\mu_B)\right>_{N_{\mathrm{max}-1}}$ 
by subtracting one from $N_\mathrm{max}$ in Eq.(\ref{trun}).
%\begin{equation}
%\sum_{B=-N_{\mathrm{max}+1}}^{N_{\mathrm{max}-1}}Z_C(B,T)e^{B\mu_B/T}.
%\end{equation}
After that, we evaluate the relative error $R_{\mathrm{ob}}(\mu_B)$ from these two expectation values;
and in this work we judge that 
the expectation value is reliable if the relative error is less than $10^{-3}$,
\begin{equation}
\label{oberror}
R_{\mathrm{ob}}(\mu_B)\equiv1-\frac{\left<O\right>_{N_{\mathrm{max}}-1}}{\left<O\right>_{N_{\mathrm{max}}}} < 10^{-3} .
\end{equation}
In this way, we can ensure that expectation values of thermodynamic observables in the baryon chemical potential region 
determined by above analysis have two significant digits at least against the truncation error.
%In fact, we can expect that reliable baryon chemical potential region is bigger than above one.  
%This is because we actually should compare results in case of $N_{\mathrm{max}}$ and $N_{\mathrm{max+1}}$
%instead of the results in case of $N_{\mathrm{max-1}}$ and $N_{\mathrm{max}}$.

\subsubsection{Thermodynamic observables}

\begin{figure}
	\center
	\includegraphics[width=0.9\hsize, bb=0 0 401 256, clip]{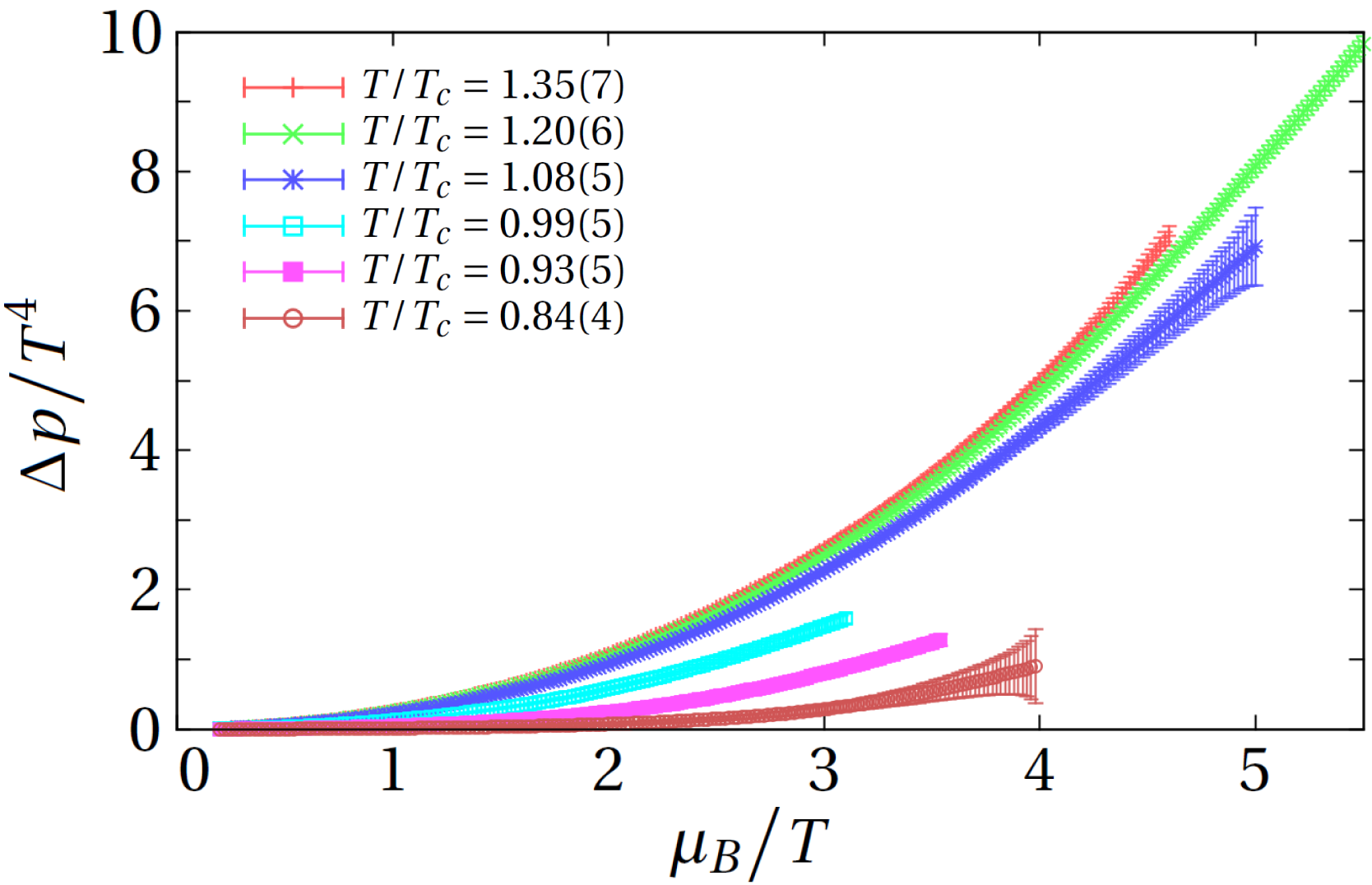}
	\caption{\label{pre-Tdep} (Color online) Chemical potential dependence of pressure.
			Red, green, blue, cyan, magenta and brown points are the results at $T/T_c=1.35, 1.20, 1.08, 0.99, 0.93 \textrm{ and } \, 0.83$.
			Upper bound of baryon chemical potential is determined by Eq.(\ref{oberror}). }
\end{figure}

Using the error estimation described in the previous subsection,
we analyze the chemical potential dependence of thermodynamic observables 
and study the validity of canonical approach.
First, we examine the pressure.
Fig.\ref{pre-Tdep} shows  
that the results of pressure above $T_c$ 
do not suffer from large error 
up to around $\mu_B/T=5$, 
and the results below $T_c$ are under control up to $\mu_B/T=3.5-4$.
On the other hand, the result just below $ T_c$ case is reliable 
only up to around $\mu_B/T=3$.
This may be because we make configurations at $\mu_0=0$ 
and the configurations are suffered 
from the fluctuation caused by the phase transition at zero density. 
We may get clearer signals if we generate configurations 
at pure imaginary chemical potential
because $T_c$ at pure imaginary chemical potential is higher than $T_c$ at zero chemical potential.

\begin{figure}
	\center
	\includegraphics[width=0.9\hsize, bb=0 0 401 256, clip]{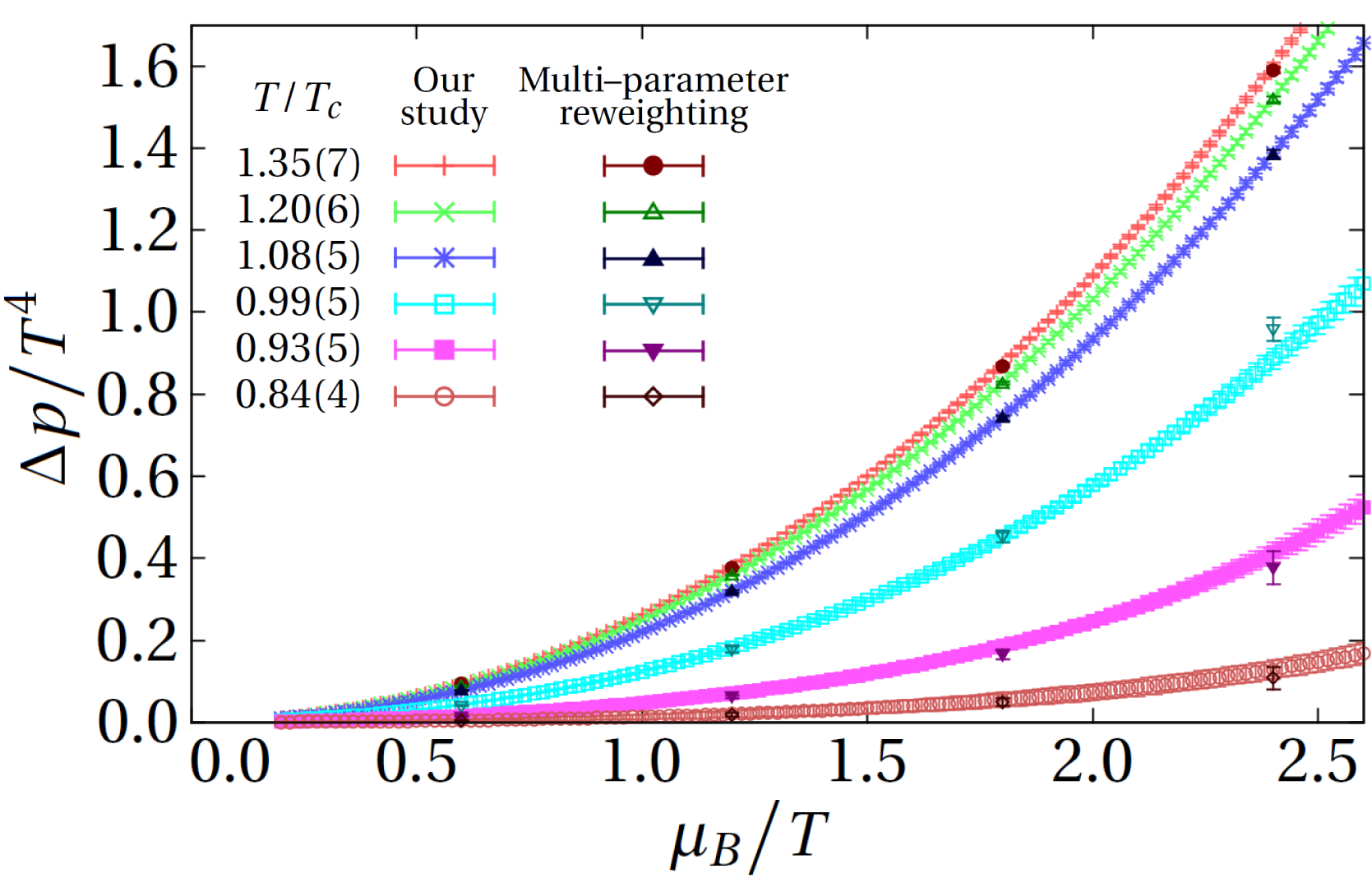}
	\caption{\label{pre-hikaku} (Color online) Comparison of pressure calculated by canonical approach and multi parameter reweighting method.
			Basically, color of data are same as Fig.\ref{pre-Tdep}. 
			Extra color: dark--red, dark--green, dark--blue, dark--cyan, dark--magenta and dark--brown points
			are the results at $T/T_c=1.35, 1.20, 1.08, 0.99, 0.93 \textrm{ and } \, 0.83$ 
			calculated by multi--parameter reweighting method. }
\end{figure}

In Fig.\ref{pre-hikaku}, we see that pressure calculated 
by canonical approach are consistent with results by MPR.

\begin{figure}
	\center
	\includegraphics[width=0.9\hsize, bb=0 0 401 256, clip]{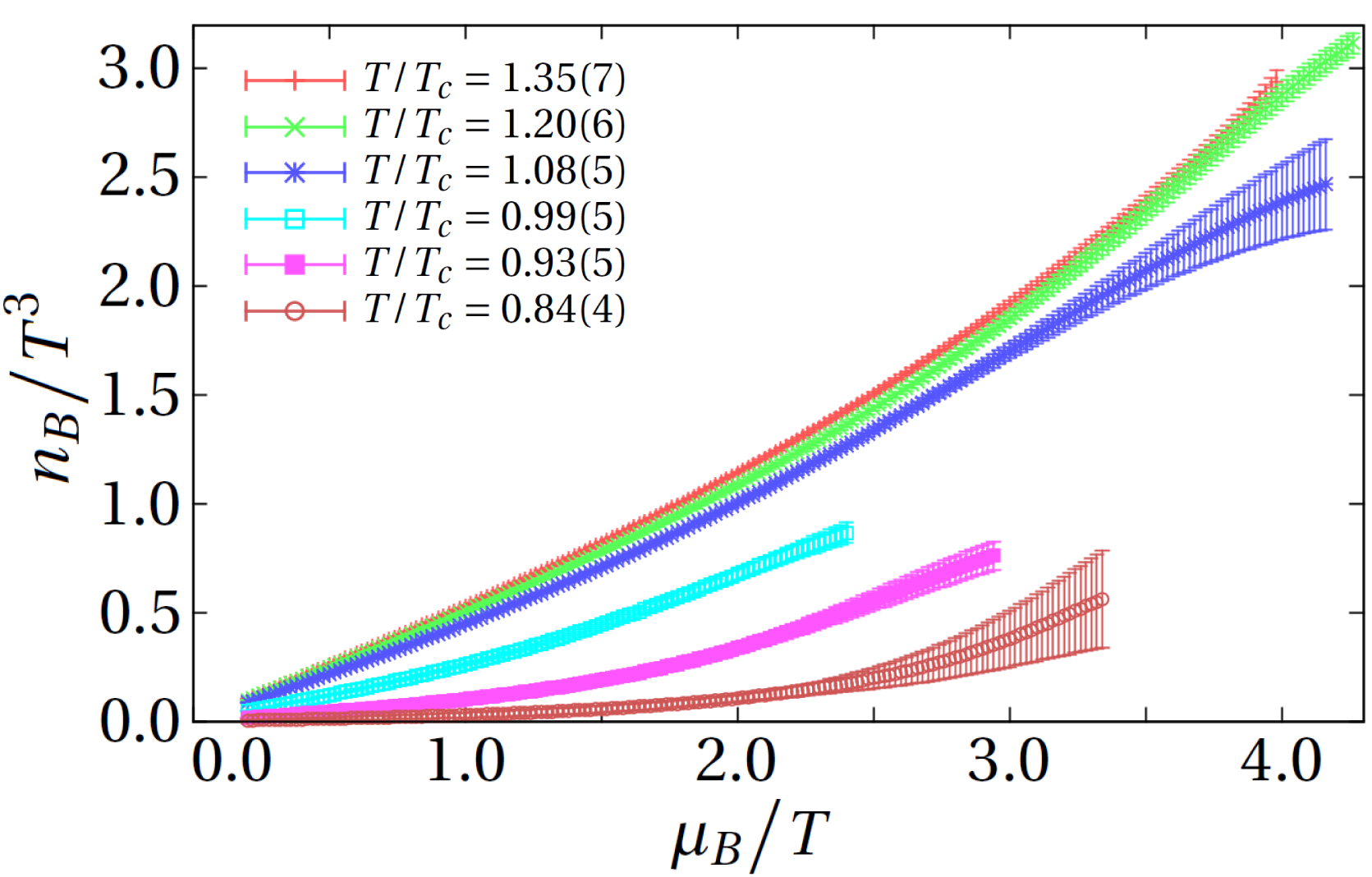}
	\caption{\label{nd-Tdep} (Color online) Chemical potential dependence of baryon number density.
			Color of data are same as Fig.\ref{pre-Tdep}. 
			Upper bound of baryon chemical potential is determined by Eq.(\ref{oberror}).}
\end{figure}

\begin{figure}
	\center
	\includegraphics[width=0.9\hsize, bb=0 0 401 256, clip]{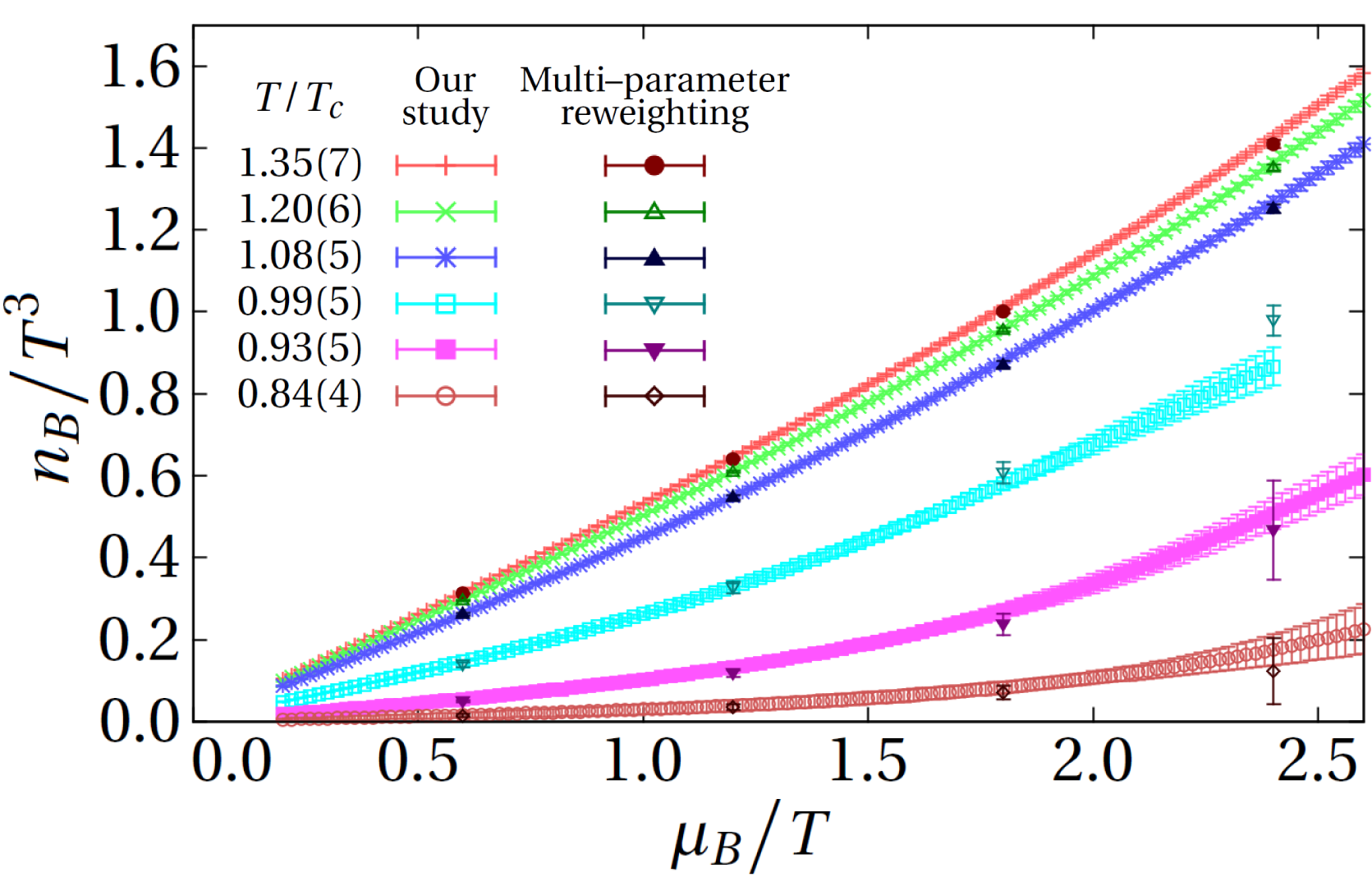}
	\caption{\label{nd-hikaku} (Color online) Comparison of the baryon number density calculated by canonical approach and multi parameter reweighting method.
			Color of data are same as Fig.\ref{pre-hikaku}. }
\end{figure}

Next, we consider the expectation value of the baryon number density.
In Fig.\ref{nd-Tdep}, we find that 
the results are reliable up to around $\mu_B/T=4$ ($\mu_B/T=3-3.5$) above $T_c$ (below $T_c$).
While, the reliable baryon chemical potential range of the result 
just below $T_c$ is limited up to $\mu_B/T=2.4$.
This may be caused by the same reason in the analysis of the pressure.

Fig.\ref{nd-hikaku} tells us that the canonical approach is consistent 
with MPR method
also in the baryon number density case.
Moreover,  we observe that the gradient of the baryon number density, $n_B$, 
as a function of baryon chemical potential
becomes smaller as the temperature decreases.
In zero temperature case, 
$n_B$ is expected to be zero
up to $\mu_B/T=m_B/T$, where $m_B$ is the lightest baryon mass of the system
and it becomes a finite value at this point.
Indeed, the data at $T/T_c=0.84$ shows such a feature.

\begin{figure}
	\center
	\includegraphics[width=0.9\hsize, bb=0 0 406 256, clip]{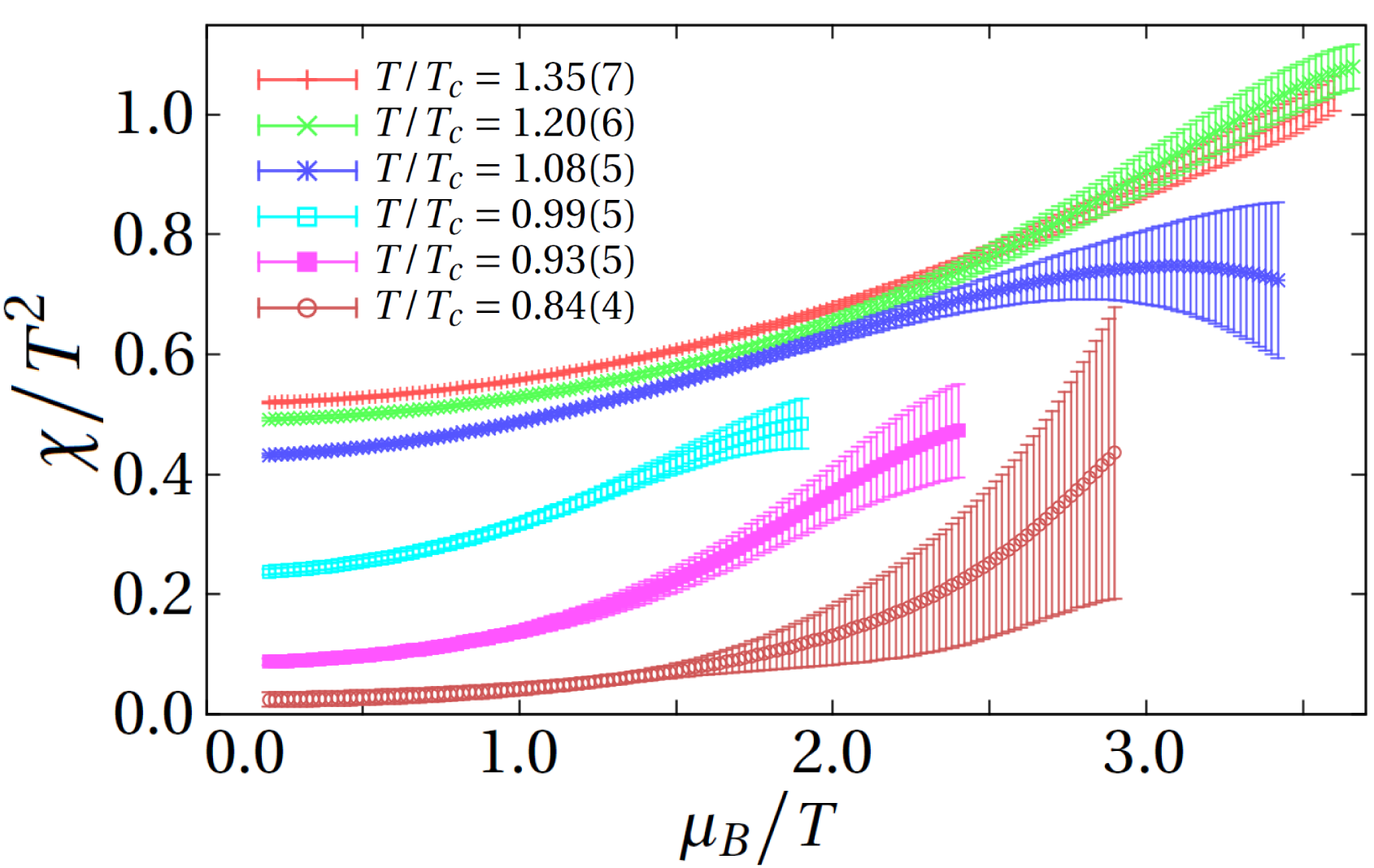}
	\caption{\label{sus-Tdep} (Color online) Chemical potential dependence of susceptibility.
			Color of data are same as Fig.\ref{pre-Tdep}. 
			Upper bound of baryon chemical potential is determined by Eq.(\ref{oberror}).}
\end{figure}

\begin{figure*}
	\center
	\includegraphics[width=0.65\hsize, bb=0 0 576 258, clip]{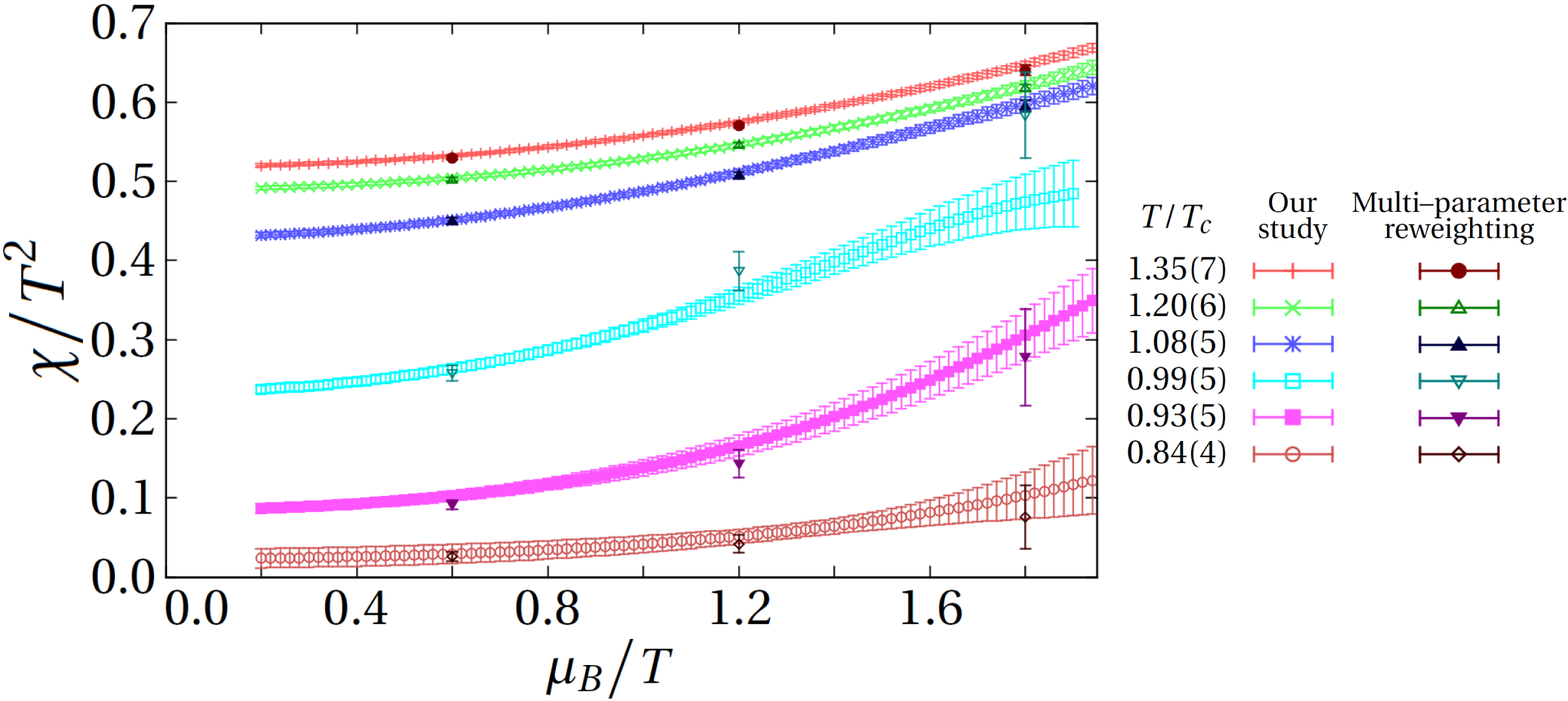}
	\caption{\label{sus-hikaku} (Color online) Comparison of susceptibility calculated by canonical approach and multi parameter reweighting method. 
			Color of data are same as Fig.\ref{pre-hikaku}. }
\end{figure*}

Finally, we investigate the susceptibility.
Fig.\ref{sus-Tdep} shows that the results above $T_c$ is reliable 
up to around $\mu_B/T=3.5$,
while the results below $ T_c$ is reliable up to $T_c=2.4-2.9$.
From Fig.\ref{sus-hikaku}, 
we find that canonical approach is very consistent 
with MPR method also in the susceptibility.
The susceptibility as a function of the $\mu_B/T$ does not show a clear peak;
we do not see yet the signal of the phase transition 
between confined phase and deconfined phase.

\section{Summary and outlook}

In this paper,
we find that the canonical approach is consistent with MPR method.
Moreover, the canonical approach provides reliable results 
beyond $\mu_B/T=3$ in almost all observables.
This is very encouraging for the first principle calculation of finite
density QCD, 
because other methods such as MPR method, Taylor expansion method and 
imaginary chemical potential method give reliable information practically 
only up to $\mu_B/T=3$.
The multiple precision calculation greatly contributes this conclusion.

The canonical approach has been investigated several times
\cite{Hasenfratz-Toussaint,deForcrand:2006ec, Li:2011ee, Alexandru:2005ix, Meng:2008hj, Li:2010qf} 
;
We brush up the method here and find that it is a useful and promising
method.
But,
we have to improve our method further 
to obtain results in more realistic condition,
i.e., lighter quark mass, large volume, finer lattice spacing and larger 
density. 
Although the hopping parameter expansion gave very interesting results
as we saw in this paper, 
the next step is to calculate the fermion determinant 
without this approximation;
we have learned in this paper that the key point is 
to calculate the determinant at imaginary chemical potential values 
that are Fourier transformed with high accuracy in Eq.(\ref{fourier}).
This requires more computational resource than the work reported here,
but within scope of the next generation high performance era.

\acknowledgments
This work is done for Zn Collaboration:  We would like to thank the members of Zn collaboration,
S. ~Sakai, A.~Suzuki and Y. ~Taniguchi for their powerful support. 
We appreciate useful discussions with K. Fukushima and Ph. de Forcrand. 
R. F. thanks the Yukawa Institute for Theoretical Physics, Kyoto University. 
Discussions during the YITP workshop YITP-T-14-03 on 
``Hadrons and Hadron Interactions in QCD'' were useful to complete this work.
R. F. also would like to thank ETH for its warm hospitality.
S. O. acknowledges T.~Eguchi for valuable discussions and encouragement.
This work is supported in part by Grants-in-Aid of the Ministry of Education (Nos. 15H03663, 26610072).
The calculations were done with SX-9 and SX-ACE at RCNP (Osaka), 
SR16000 at  the Yukawa Institute for Theoretical Physics (Kyoto).

\end{document}